\documentclass{aa}

\usepackage{graphicx}
\usepackage{txfonts}
\usepackage{threeparttable}
\usepackage{multirow}
\usepackage{CJK}
\usepackage{color}

\begin{document}

\begin{CJK*}{GB}{gbsn} 

   \title{The extraneous eclipses on binary light curves: \\KIC 5255552, KIC 10091110, and KIC 11495766}

   \subtitle{}

   \author{J. Zhang (张嘉)
          \inst{1,2}
          \and
          S. B. Qian (钱声帮)\inst{1,2}
          \and
          S. M. Wang (王树民)\inst{3}
          \and
          L. L. Sun (孙磊磊)\inst{1,2}
          \and
          Y. Wu (吴悦)\inst{4}
          \and
          L. Q. Jiang$^*$ (江林巧)\inst{5}
          }

   \institute{Yunnan Observatories, Chinese Academy of Sciences, Kunming 650216, China\\
              \email{zhangjia@ynao.ac.cn}
         \and
             Key Laboratory of the Structure and Evolution of Celestial Objects, Chinese Academy of Sciences, Kunming 650216, China
         \and
             School of Mathematics and Physics, Handan University, Handan 056005, China
         \and
             Key Laboratory of Optical Astronomy, National Astronomical Observatories, Chinese Academy of Sciences, Beijing 100012, China
         \and
             School of Physics and Electronic Engineering, Sichuan University of Science ＆ Engineering, Zigong 643000, China
             \email{jianglinqiao11@163.com}
             }

   \date{Received September 15, 1996; accepted March 16, 1997}

  \titlerunning{The extraneous eclipses on binary light curves}
  \authorrunning{J. Zhang et al.}


  \abstract
   {}
   {We aim to find more eclipsing multiple systems and obtain their parameters, thus increasing our understanding of multiple systems.}
   {The extraneous eclipses on the \textit{kepler} binary light curves indicating extraneous bodies were searched. The binary light curves were analyzed using the binary model, and the extraneous eclipses were studied on their periodicity and shape changes.}
   {Three binaries with extraneous eclipses on the binary light curves were found and studied based on the \textit{Kepler} observations. The object KIC 5255552 is an eclipsing triple system with a fast changing inner binary and an outer companion uncovered by three groups of extraneous eclipses of $862.1(\pm0.1)$ d period. The KIC 10091110 is suggested to be a double eclipsing binary system with several possible extraordinary coincidences: the two binaries share similar extremely small mass ratios ($0.060(13)$ and $0.0564(18)$), similar mean primary densities ($0.3264(42)\;\rho_\odot$ and $0.3019(28)\;\rho_\odot$), and, most notably, the ratio of the two binaries' periods is very close to integer 2 (8.5303353/4.2185174 = 2.022). The KIC 11495766 is a probable triple system with a $\sim120.73$ d period binary and (at least) one non-eclipse companion. Furthermore, very close to it in the celestial sphere, there is a blended background stellar binary of 8.3404432 d period. A first list of 25 eclipsing multiple candidates is presented, with the hope that it will be beneficial for study of eclipsing multiples.}
   {}


   \maketitle
%

\section{Introduction} \label{sec:intro}

The discovery for an extraneous component in a previous known stellar system by finding extraneous eclipses on the light curves has a long history. The earliest binary star system $\beta$ Persei\footnote{Actually a triple star system. It is also the first discovered variable star. Its periodic light changes were recorded before 1163 B.C. by ancient Egyptians (Jetsu et al. 2013; Jetsu \& Porceddu 2015). The unrecorded discovery may be even earlier as it is the most easily discovered variable to the naked eye.}, known as Algol, was discovered in this way more than 3000 years ago. Most of the binary systems were found (and also studied) by the eclipse signals, hence the name `eclipsing binaries'. This method also works for the multiple systems. The extraneous eclipses on the light curves of previous known binaries suggest a new component accompanying the binary system, just like the eclipses on a single star's light curve suggest a companion to the single star. Since the studies on eclipsing binaries have provided numerous fundamental properties for the stellar field, similarly, the eclipsing multiple systems will also provide important information, especially in regards to the star formation environment.

The biggest problem in the study of eclipsing multiple systems is small sample sizes, and this situation should not be caused by the scarcity of multiple systems but caused by the limitations of observation. The emergence odds of the extraneous eclipses are slim, and therefore until the release of the  \textit{Kepler} space telescope it has been hard to find them confidently. The \textit{Kepler} telescope is devoted to discovering earth-size planets orbiting other stars, and so to monitor photometrically many kinds of stars including binaries. The extraordinary features of the \textit{Kepler} observations, that is, the long-term (four years) uninterrupted brightness monitoring in a 115 $deg^2$ fixed field of view with unprecedented photometric precision (20 per million) and high limiting magnitude (21st magnitude), are ideal and suitable for searching for extraneous eclipses on the binary light curves. In fact, the \textit{Kepler} telescope was designed and operated for the purpose of finding extraneous eclipses indicating the planets around stars.

In this paper, three binaries with extraneous eclipses, \object{KIC 5255552}, \object{KIC 10091110} and \object{KIC 11495766}, are studied. Section 2 illustrates the data reduction. Studies of the three objects are outlined in Section 3. A first list of eclipsing multiple candidates is presented in Section 4. We summarize our work in the last section.

\section{Data reduction}

Binary light curves were provided by the \textit{Kepler} mission, and they were downloaded from Mikulski Archive for Space Telescopes (MAST) at a convenient individual webpage\footnote{\url{http://archive.stsci.edu/pub/kepler/lightcurves/tarfiles/EclipsingBinaries/}} on October 27, 2015. Besides the photometric data, the Kepler Eclipsing Binary Catalog (Pr{\v s}a et al. 2011; Slawson et al. 2011; Matijevi{\v c} et al. 2012; Conroy et al. 2014; Kirk et al. 2016) is another important resource. The catalog provides multiple orbital periods for some binaries, which led us to the extraneous eclipses in these binaries.

\subsection{ Normalization of Kepler light curves}

The Pre-Search Data Conditioning (PDC) fluxes in the Kepler data were selected `to identify and correct flux discontinuities that cannot be attributed to known spacecraft or data anomalies' (Kepler Data Processing Handbook\footnote{\url{http://archive.stsci.edu/kepler/manuals/KSCI-19081-001_Data_Processing_Handbook.pdf}}; Fraquelli et al. (2014)). However, there are still unphysical jumps and long-term (week or longer) variations in the PDC fluxes. Almost all the jumps occurred after the light curve gaps (where there is no datum); about half of the gaps were the borders of the observation quarters of the \textit{Kepler} spacecraft, and the other half were caused by the missing data.

These unphysical jumps and long-term variations can be eliminated simultaneously by the flux normalization on each light curve segment (divided by the gaps). However, the long-term variations contain not only the unphysical but also the physical parts, and the physical parts will be also eliminated by the flux normalization together with the unphysical parts. Fortunately, the elimination of the long-term variations does not affect the analysis on the binary light variations, because the binary light variations are always in short term, whether the orbital period is long or short (for the long-period binaries, their light variations, that is, the eclipses, are also very short). In fact, the elimination is often more conducive to the analysis of binary light curves, because some irrelevant light variations (such as magnetic activities, pulsations) are removed.

The flux normalization steps are: (1) the light curves are divided into dozens of segments depending on the width of the gaps on the light curves. If the width of a gap is larger than one day (according to our experiences, somewhat arbitrary), the light curve will be divided into two segments at that gap; (2) each light segment is fitted by a cubic polynomial model, and their fluxes are divided by the fitting values to be normalized. In the fitting, the large scattered points are removed to get a better fit in an iterative process (up to ten times practically); and (3) all the normalized segments are stitched together to form the whole normalized light curve.

\subsection{Data binning}\label{sec:phase_cal}

Before the light curve analysis, the barycentric julian dates (BJD) need to be transformed to phase. The orbital periods for calculating the phases were taken from the Kepler Eclipsing Binary Catalog, but the reference minimum times were re-measured, not because the values in the catalog are not accurate enough, but because the mean reference minimum times\footnote{The mean reference minimum time is calculated in the following way: 1, calculate the phase and fold them to phase 0.5 to 1.5 with an initial reference minimum time. 2, fit the minimum at phase 1 to obtain its precise minimum phase. 3, the difference between this phase value and 1 is used to correct the reference minimum time. 4, iterate these previous three steps until the difference is small enough.} (instead of a single minimum time) are needed. After the phase calculation, the dense phase points are binned to reduce the data amount, and also to get rid of the large scattered points.

The binning phase data points were made to be not evenly distributed along the phases but inclined to place more points at eclipses. In other words, the points at the eclipsing parts will be more dense along the abscissa. It is believed that this kind of data binning is more suitable for the binary analysis, because more points (so more weight) should be given to the important eclipsing parts. The final binning phase data are analyzed in the following section.

\section{Three \textit{Kepler} binaries with extraneous eclipses}

\subsection{KIC 5255552}

\subsubsection{The extraneous eclipses on KIC 5255552}
The object KIC 5255552 was discovered and analyzed as an eclipsing binary with 32.448635 d period by Pr{\v s}a et al. (2011) and Slawson et al. (2011) based on the \textit{Kepler} data. Borkovits et al. (2015, 2016) measured and analyzed its $O-C$ curve with a new and robust tool which is capable of dealing with eccentric binaries, and determined the reliable parameters. Borkovits et al. (2016) obtained that the 32.448635 d period inner binary has a 0.7($\pm0.1$)$M_{\odot}$ tertiary body revolving around it with 862.1($\pm0.1$) d period at only 6.4($\pm0.1$) degrees of mutual inclination, and also noted three outer eclipsing events (Borkovits et al. 2015).

\begin{figure*}
\centering
\includegraphics[scale=.45]{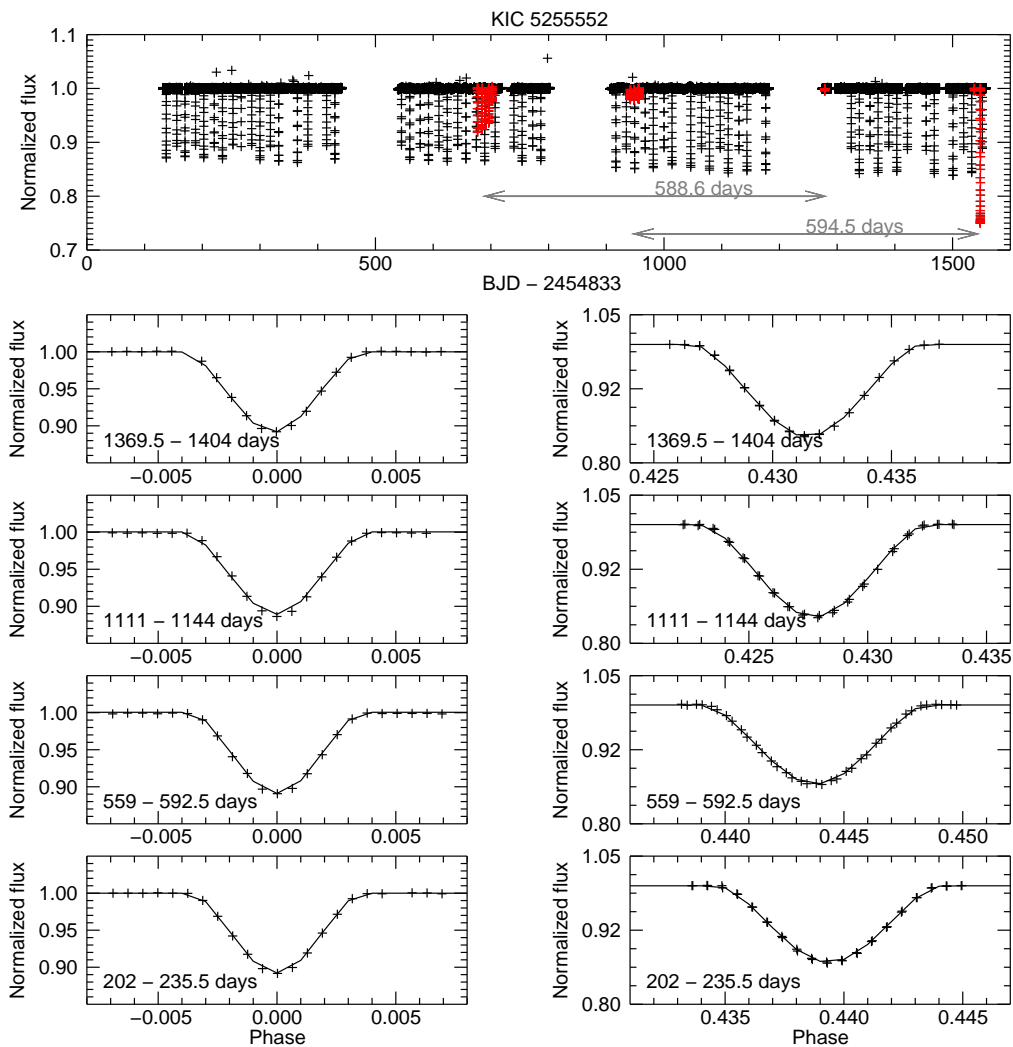}
\caption{The light curves of KIC 5255552. Upper: The whole light curves. The red points are the extraneous eclipses that can been seen more clearly in Figure \ref{KIC005255552_extra_eclipses.eps}, and the gray arrows are used to illustrate the time intervals between the extraneous eclipses. Lower eight panels: The four selected light curve parts in phase form, showing the eclipses at phase zero (left) and phase around 0.43 (right) with their fittings (black line). \label{KIC005255552.eps}}
\end{figure*}

\begin{figure*}
\centering
\includegraphics[scale=.45]{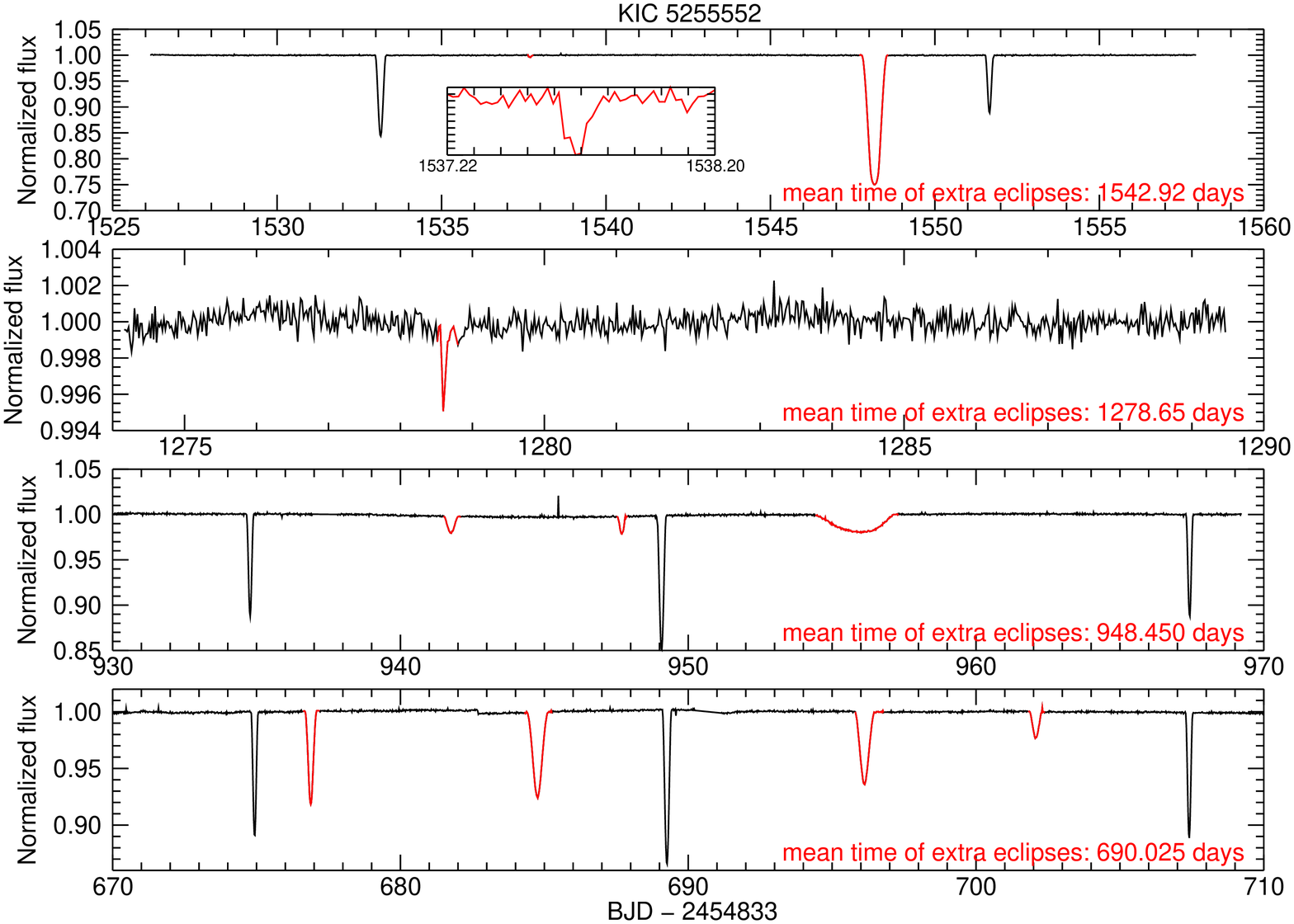}
\caption{The extraneous eclipses of KIC 5255552. The red points in all the panels are the extraneous eclipses. The mean times of the extraneous eclipses for each group are marked in each panel. In the top panel, a small additional panel is used to magnify a tiny extraneous eclipse around 1537.7 d. \label{KIC005255552_extra_eclipses.eps}}
\end{figure*}

In this paper, four groups of extraneous eclipses are reported, illustrated in Figures \ref{KIC005255552.eps} and \ref{KIC005255552_extra_eclipses.eps}. The extraneous eclipses, emerging in groups, are generated by the mutual blocking between the binary components and the outer companion. When the outer companion passes in front of (or behind) the binary, it will block (or be blocked by) the two components of the binary. The blocking may generate several eclipses for one passage, as shown in Figure \ref{KIC005255552_extra_eclipses.eps}.

The first, second, and fourth groups of extraneous eclipses (see the red points in Figure \ref{KIC005255552.eps}) are consistent with the 862.1($\pm0.1$) d period of the third companion concluded by Borkovits et al. (2016). These two mutual confirmations are strong evidence for the triplicity of this object. However, a much shallower and narrower fading was also found at $\sim1279$ d, which happens to constitute a period of $\sim$589 d with another three groups (see the gray arrows in Figure \ref{KIC005255552.eps}). Considering that the $\sim1279$ fading is not analogous to any other three groups, so the apparent $\sim$589 d period may simply be a coincidence rather than a real period. If this small fading is physical, it indicates another new companion that makes the KIC 5255552 a quadruple system.

\subsubsection{The light analysis of the inner binary}

The inner 32.448635 d period binary was changing dramatically on eclipses (see the minimum position changes in lower-right panels of Figure \ref{KIC005255552.eps}), so the analysis on the whole light curves is unfeasible. In view of this, four light curve parts with lengths of about one period were selected and analyzed independently. The results are listed in Table \ref{tab:KIC005255552_solution} and shown in Figure \ref{KIC005255552.eps}\footnote{The light analysis was carried out by setting the shallower eclipses fixed at phase zero, and so the position of the deeper eclipses vary with time.}.

\begin{table*}
\small
\begin{center}
\begin{threeparttable}
\caption{The binary light curve solutions of KIC 5255552. \label{tab:KIC005255552_solution}}
\begin{tabular}{lllll}
\hline
\hline
\multirow{2}{*}{Parameters} &  \multicolumn{4}{c}{Time span (BJD - 2454833)}\\
                                                                             &    202 - 235.5 d            &   559 - 592.5 d       &   1111 - 1144 d       &  1369.5 - 1404 d        \\
\hline
Period (d)                                                                   &    32.4676(7)               &   32.4407(8)          &   32.4668(10)         &  32.4473(3)             \\
Mode                                                                         &    detached binary          &   detached binary     &   detached binary     &  detached binary        \\
Orbital eccentricity $e$                                                     &    0.21160(81)              &   0.2043(12)          &   0.1868(12)          &  0.19960(70)            \\
Argument of periasrtron of star 2 $\omega$                                   &    2.0291(20)               &   2.0099(28)          &   2.2164(53)          &  2.1311(23)             \\
Orbital inclination $i[^{\circ}]$                                            &    89.1580(22)              &   89.1700(91)         &   89.1540(23)         &  89.1770(24)            \\
Mass ratio $m_2/m_1$                                                         &    0.5 (fixed)              &   0.5 (fixed)         &   0.5 (fixed)         &  0.5 (fixed)            \\
Primary temperature $T_2$\tnote{a}                                           &    5125 (fixed)             &   5125 (fixed)        &   5125 (fixed)        &  5125 (fixed)           \\
temperature ratio $T_1/T_2$                                                  &    1.1731(16)               &   1.1725(49)          &   1.1840(22)          &  1.2020(23)             \\
Luminosity ratio $L_1/(L_1 + L_2)$ in band Kepler                            &    0.84708(75)              &   0.8440(27)          &   0.8267(13)          &  0.8384(15)             \\
Luminosity ratio $L_2/(L_1 + L_2)$ in band Kepler                            &    0.15292(75)              &   0.1560(27)          &   0.1733(13)          &  0.1616(15)             \\
Luminosity ratio $L_3/(L_1 + L_2 + L_3)$ in band Kepler                      &    0.15001(11)              &   0.14991(41)         &   0.14993(21)         &  0.15001(23)            \\
Modified dimensionless surface potential of star 1 $\Omega_1$                &    109.05(35)               &   109.72(67)          &   110.01(47)          &  110.55(47)             \\
Modified dimensionless surface potential of star 2 $\Omega_2$                &    91.22(23)                &   90.8(13)            &   83.95(19)           &  85.30(51)              \\
Radius of star 1 (relative to semimajor axis) in pole direction              &    0.018498(60)             &   0.01835(11)         &   0.018303(79)        &  0.018232(78)           \\
Radius of star 2 (relative to semimajor axis) in pole direction              &    0.011262(29)             &   0.01136(17)         &   0.012263(29)        &  0.012043(74)           \\
Radius of star 1 (relative to semimajor axis) in point direction             &    0.018499(60)             &   0.01835(11)         &   0.018303(79)        &  0.018232(78)           \\
Radius of star 2 (relative to semimajor axis) in point direction             &    0.011262(29)             &   0.01136(17)         &   0.012264(29)        &  0.012043(74)           \\
Radius of star 1 (relative to semimajor axis) in side direction              &    0.018498(60)             &   0.01835(11)         &   0.018303(79)        &  0.018232(78)           \\
Radius of star 2 (relative to semimajor axis) in side direction              &    0.011262(29)             &   0.01136(17)         &   0.012264(29)        &  0.012043(74)           \\
Radius of star 1 (relative to semimajor axis) in back direction              &    0.018499(60)             &   0.01835(11)         &   0.018303(79)        &  0.018232(78)           \\
Radius of star 2 (relative to semimajor axis) in back direction              &    0.011262(29)             &   0.01136(17)         &   0.012264(29)        &  0.012043(74)           \\
Equal-volume radius of star 1 (relative to semimajor axis) $R_2$             &    0.018471(32)             &   0.018356(59)        &   0.018302(42)        &  0.018216(41)           \\
Equal-volume radius of star 2 (relative to semimajor axis) $R_1$             &    0.011278(15)             &   0.011329(92)        &   0.012271(15)        &  0.012078(39)           \\
Radius ratio $R_1/R_2$                                                       &    1.6378(36)               &   1.620(14)           &   1.4915(39)          &  1.5082(60)             \\
The density of star 1 (relative to $\rho_\odot=1410.04084 Kg/m^3$) $\rho_1$  &    1.3465(92)               &   1.372(18)           &   1.384(13)           &  1.404(13)              \\
The density of star 2 (relative to $\rho_\odot=1410.04084 Kg/m^3$) $\rho_2$  &    2.9578(80)               &   2.918(47)           &   2.2961(57)          &  2.408(16)              \\
\hline
\hline
\end{tabular}
\begin{tablenotes}
  \item[a]The temperature was taken from Rowe et al. (2015).
\end{tablenotes}
\end{threeparttable}
\end{center}
\end{table*}

The mass ratio was taken from Borkovits et al. (2015)\footnote{
Unlike the case of circular orbit, in an eccentric orbit like KIC 5255552, the deeper eclipse is not necessary generated by the massive component being blocked by its companion. It is possible that the deeper eclipse is generated by that of the less massive component. So this need to be clarified first: Based on the time of periastron passage and the argument of periastron of the less massive component provided by Borkovits et al. (2015), it can be worked out that for KIC 5255552, the deeper eclipse corresponds to the massive component being blocked by its companion.}, and was fixed in the fitting process. This is because the photometric determination on the mass ratio for a long-period detached binary tends to have considerably great uncertainty. The periods in Table \ref{tab:KIC005255552_solution} were calculated separately by subtracting two successive deeper minimum times within the selected time span, and they differ somewhat, which can be seen numerically. Similarly, the other parameters for different time spans have changes of a few percent.

The changing parameters with time indicate that the orbit of the 32.448635 d period binary may differ from Keplerian orbit by some degree. This is likely to be due to the disturbance by the massive ($M_{outer}/M_{inner}=0.29$), close ($P_{outer}/P_{inner}=26.2$), and coplanar ($i_m=6.4$) outer companions (Borkovits et al. 2016). The KIC 5255552 is not the only system with a non-Keplerian orbit, the object KIC 2856960 is a known triple system whose extraneous eclipses cannot be fitted while satisfying the Kepler's laws (Marsh et al. 2014).

In summary, the object KIC 5255552 is a hierarchical eclipsing multiple system consisting of three (or four) component stars. The orbit of the inner binary is changing visibly with time, which makes this object a good sample for studying the interaction in the multiple systems.

\subsection{KIC 10091110}

\subsubsection{Light curve analysis}
\begin{figure*}
\centering
\includegraphics[scale=.45]{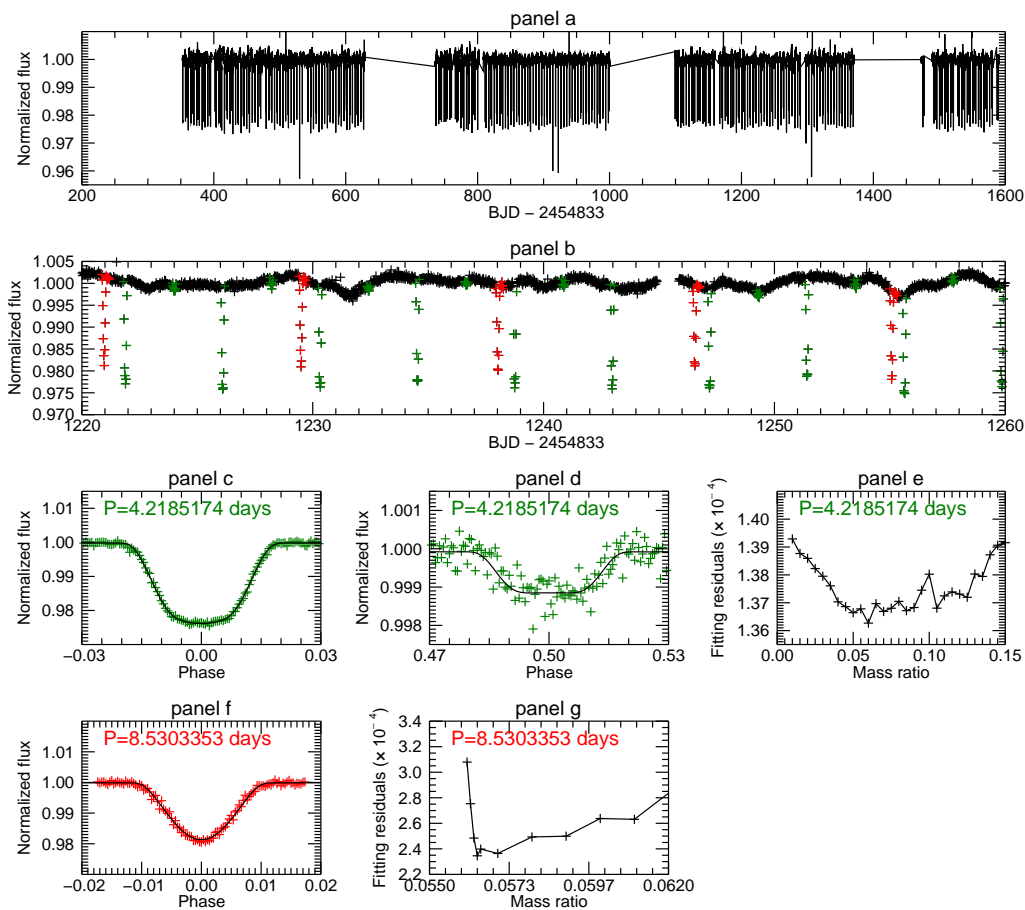}
\caption{The light curves of KIC 10091110 with the solutions. Panel a: The whole light curves of KIC 10091110. Five large scattered points are not shown in the visible region. Panel b: A light curve part from 1220 d to 1260 d. The green and red points are the eclipses of 4.2185174 d and 8.5303353 d period, respectively. Panel c and d: The folded and binned phase curves by 4.2185174 d period around the primary and secondary minimums with the black fitting lines. Panel e: The mass ratio search curve of 4.2185174 d period binary. Panel f and g: The same as panel c and e but for 8.5303353 d period. \label{KIC010091110.eps}}
\end{figure*}

The object KIC 10091110 has two sets of very obvious eclipses with periods of 4.2185174 d and 8.5303353 d, which are shown in Figure \ref{KIC010091110.eps}, panels (a) and (b)\footnote{It should be noted that, in panel (a), the deep eclipses at around 530 d, 920 d and 1300 d are not an independent set of eclipses, but are only the superposition of the two sets of eclipses.}. The 4.2185174 d period eclipses have secondary minimums that can be explicitly seen in panel (d), and the light curves can be well fitted by the binary model Wilson-Devinney program (Wilson \& Devinney 1971; Wilson 1979, 1990; Van Hamme \& Wilson 2007; Wilson 2008; Wilson et al. 2010; Wilson 2012). The atmosphere parameters of KIC 10091110 were $T_{eff}=6185(204)$, $\log g=4.472(0.166)$ and $[Fe/H]=-0.18(0.27)$ (Rowe et al. 2015), and they are employed in the light analysis. The fits are shown in panels (c) and (d), and the parameters are listed in Table \ref{tab:KIC010091110_solution}.

Like the 4.2185174 d period eclipses, the 8.5303353 d period eclipses were analyzed with the Wilson-Devinney program. Due to the lack of secondary eclipse (no secondary eclipse was found after the data folding and binning), its solution bears much more uncertainty. In view of this point, the 8.5303353 d period eclipses were also analyzed by the Spot and Transit Modeling Tool (STMT, Sun et al. 2017), and the results are listed in the last column of Table \ref{tab:KIC010091110_solution} for comparison. The comparison shows that,  generally, only the primary mean densities coincide with each other, and the differences of the other parameters reach 50\%. For parameters concerned with the secondary star, such as luminosity ratio and radius ratio, their uncertainties are very large.

\begin{table*}
\small
\begin{center}
\begin{threeparttable}
\caption{The binary light curve solutions of KIC 10091110. \label{tab:KIC010091110_solution}}
\begin{tabular}{llll}
\hline
\hline
\multirow{2}{*}{Parameters} &  \multicolumn{3}{c}{KIC 10091110}\\
                                                                        &       WD program         &     WD program        &   STMT program\tnote{a}         \\
\hline
Period (days)                                                           &       4.2185174          &     8.5303353         &   8.5303353                     \\
mode                                                                    &       detached binary    &     detached binary   &   detached binary               \\
orbital inclination $i[^{\circ}]$                                       &       84.299(43)         &     81.06(15)         &   83.8949(3)                    \\
mass ratio $m_2/m_1$                                                    &       0.060(13)          &     0.0564(18)        &   ---                           \\
primary temperature $T_1$\tnote{b}                                      &       6185 (fixed)       &     6185 (fixed)      &   6183(188)                     \\
temperature ratio $T_2/T_1$                                             &       0.5685(48)         &     0.297(77)         &   0.207(9)                      \\
Luminosity ratio $L_1/(L_1 + L_2)$ in band Kepler                       &       0.9990381(47)      &     0.9997082(35)     &   ---                           \\
Luminosity ratio $L_2/(L_1 + L_2)$ in band Kepler                       &       0.0009619(47)      &     0.0002918(35)     &   ---                           \\
Luminosity ratio $L_3/(L_1 + L_2 + L_3)$ in band Kepler                 &       0.0065(48)         &     0.023(11)         &   ---                           \\
Modified dimensionless surface potential of star 1 $\Omega_1$           &       7.779(34)          &     12.06(10)         &   ---                           \\
Modified dimensionless surface potential of star 2 $\Omega_1$           &       4.44(63)           &     2.132(30)         &   ---                           \\
Radius of star 1 (relative to semimajor axis) in pole direction         &       0.12955(52)        &     0.08304(72)       &   ---                           \\
Radius of star 2 (relative to semimajor axis) in pole direction         &       0.0202(55)         &     0.0843(73)        &   ---                           \\
Radius of star 1 (relative to semimajor axis) in point direction        &       0.12973(53)        &     0.08307(72)       &   ---                           \\
Radius of star 2 (relative to semimajor axis) in point direction        &       0.0202(56)         &     0.0863(81)        &   ---                           \\
Radius of star 1 (relative to semimajor axis) in side direction         &       0.12970(53)        &     0.08307(72)       &   ---                           \\
Radius of star 2 (relative to semimajor axis) in side direction         &       0.0202(56)         &     0.0848(74)        &   ---                           \\
Radius of star 1 (relative to semimajor axis) in back direction         &       0.12972(53)        &     0.08307(72)       &   ---                           \\
Radius of star 2 (relative to semimajor axis) in back direction         &       0.0202(56)         &     0.0861(80)        &   ---                           \\
Equal-volume radius of star 1 (relative to semimajor axis) $R_1$        &       0.12965(28)        &     0.08331(38)       &   ---                           \\
Equal-volume radius of star 2 (relative to semimajor axis) $R_2$        &       0.0202(29)         &     0.0858(40)        &   ---                           \\
Sum of the Fractional Radii $R_1+R_2$                                   &       0.1499(29)         &     0.1691(40)        &    0.12240(1)                   \\
Radius ratio $R_2/R_1$                                                  &       0.156(23)          &     1.030(48)         &    0.42478(3)                   \\
The density of star (relative to $\rho_\odot=1410.04084 Kg/m^3$) 1 $\rho_1$ &       0.3264(42)         &     0.3019(28)        &    0.29058(7)                   \\
The density of star (relative to $\rho_\odot=1410.04084 Kg/m^3$) 2 $\rho_2$ &       5.163(90)          &     0.015590(81)      &   ---                           \\
Mass of star 1 ($M_\odot$)                                              &       ---                &     ---               &   1.20(5)                       \\
Radius of star 1 ($R_\odot$)                                            &       ---                &     ---               &   1.60(2)                       \\
\hline
\hline
\end{tabular}
\begin{tablenotes}
  \item[a]Spot and Transit Modeling Tool (Sun et al. 2017)
  \item[b]The temperature was taken from Rowe et al. (2015).
\end{tablenotes}
\end{threeparttable}
\end{center}
\end{table*}

\subsubsection{Orbital structure of KIC 10091110}

Based on the available information including the light curve solutions above, we now wish to inquire into the orbital structure of KIC 10091110. Is KIC 10091110 an eclipsing triple system, or a system consisting of two irrelevant eclipsing binaries? Or what is the origin of the two sets of eclipses? There are three weak pieces of evidence hinting at an eclipsing triple system: (1) The third light proportions in the light curve solutions of two periods are both marginal (0.0065(48) and 0.023(11)), indicating only one luminous primary star with two faint companions. (2) The mean density figures of star 1 for the two periods are coincidentally close to each other (0.3264 is close to 0.3019 or 0.29058), indicating the primary stars of the two periods may be the same one. (3) The ratio of the two periods is very close to integer 2, that is 8.5303353/4.2185174 = 2.022. This suggests that the two orbits may be in mean-motion resonance, thus belonging to the same triple system. However, these observations are not enough to confirm the triple system structure; the third light proportion is not reliable in the light curve solution, and the same primary star density and the integer period ratio may simply be coincidences.

In the meantime, there is strong evidence, in two separate forms, against the triple system scenario: (1) No decent eclipse timing variation (ETV) was found from the two sets of eclipses. Our measurements show that the amplitudes of ETV (if there is one) will be no more than 0.008 d (for 4.2 d periods) and 0.005 d (for 8.5 d periods). Furthermore, no ETV was reported from the careful sweeping work on the entire \textit{Kepler} binaries performed by Rappaport et al. (2013) and Borkovits et al. (2015, 2016) who are able to discover binaries with ETV amplitudes as small as 0.0002 d. The absence of measurable ETV is extremely unlikely for a triple system with a period ratio of only 2. If considering the 2:1 mean-motion resonance, the ETV amplitude of 4.2185174 d period (by adopting the parameters of Table \ref{tab:KIC010091110_solution}) may reach a maximum of 0.5 d (by Eq. 33 in Agol et al. (2005)) that significantly exceeds the detection limit. (2) A triple system with a period ratio of 2:1 cannot be stable. Equation 1 in Chambers et al. (1996) was used to calculate the stability based on the parameters of Table \ref{tab:KIC010091110_solution}. The result shows that the triple structure is far from stable. Compared to the three pieces of evidence supporting the triple structure, these observations to the contrary are much stronger. Therefore, the two sets of eclipses cannot be generated from an eclipsing triple system, and the object KIC 10091110 cannot be a triple system.

Since KIC 10091110 cannot be a triple system, could it be a double binary system? It probably is. The evidence mentioned above is reexamined for the double binary scenario. The absence of ETV (as well as the changes in eclipse depth) and the system stability are consistent with a double binary structure. However, the tiny proportion of the third light on both binary light solutions goes against a double binary structure. Because the third light of one binary is the light of the other binary, the third light cannot be small for the two binaries. A reasonable case is that one third light proportion is small, and the other is large. Considering that the third light is not a reliable parameter in the light solution, especially for the 8.5 d period binary solution (due to the absence of secondary eclipse mentioned above), this is not strong evidence against the double binary structure. As for the close mean density of primary stars and the integer period ratio, although these two points do not contradict the double binary structure, they are amazing coincidences.

In summary, we tend to think that KIC 10091110 is a double binary system but with extraordinary coincidences: the two binaries show huge differences in luminosity, have primary stars with close mean density, the ratio of the two binaries' periods is close to integer 2, and both have extremely small mass ratio. Whether these coincidences suggest some physical relationship between the two pairs of binaries or whether they belong to one gravitationally bound system is unclear. Further, do they originate in the same environment? Otherwise, why so many coincidences?

\subsection{KIC 11495766}

\subsubsection{The 8.3404432 d period eclipses}
\begin{figure*}
\centering
\includegraphics[scale=.45]{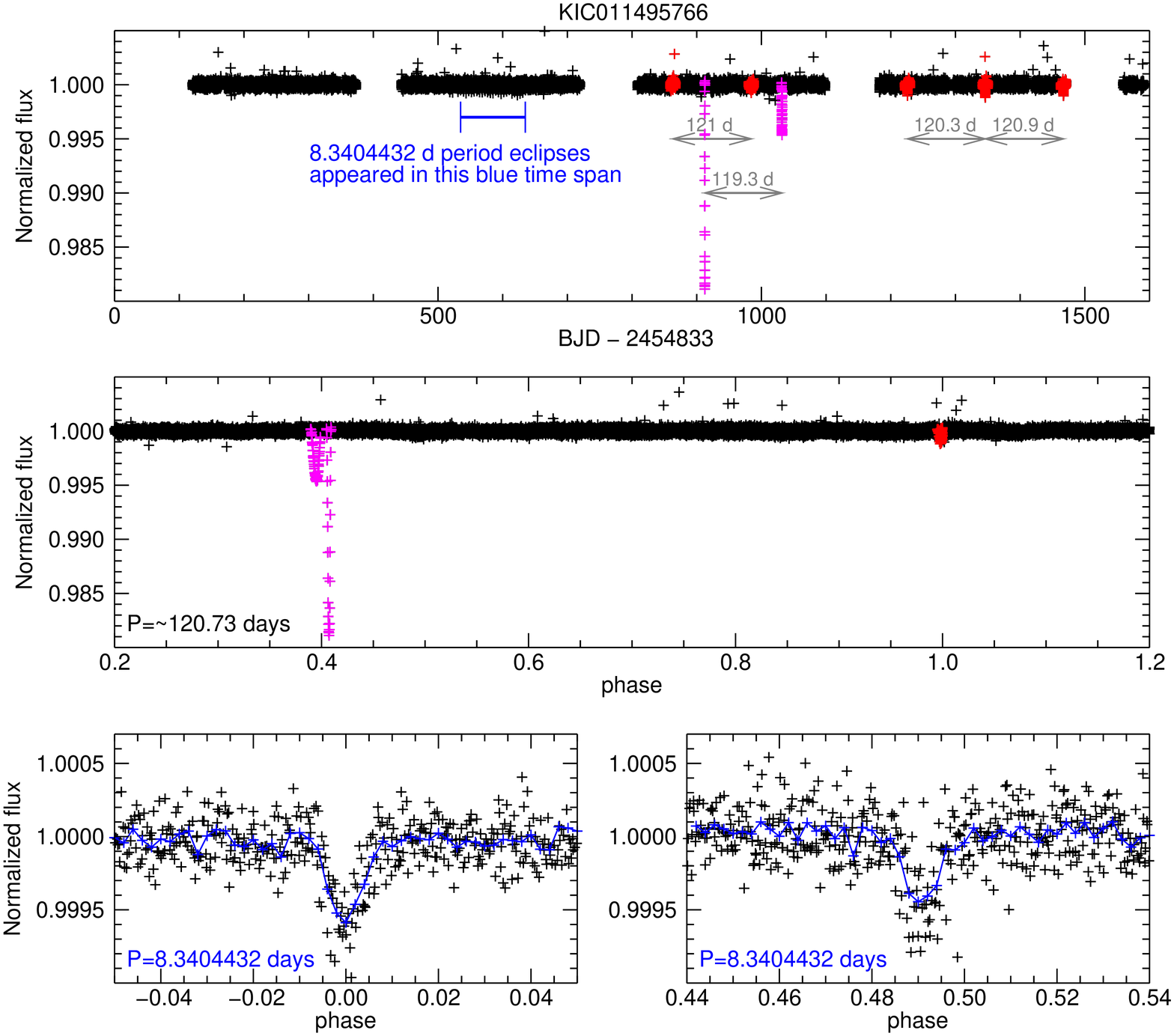}
\caption{Upper: The whole light curve of KIC 11495766. The blue segment line indicates the time span when the 8.3404432 d period eclipses show up, and the red and magenta points represent the extraneous eclipses. The gray arrows with values indicate the time intervals between the eclipses. Middle: The folded phase curves of the eclipses of $120.73$ d period. The red and magenta points correspond to those in the upper panel. Lower panels: The folded phase curves of 8.3404432 d period eclipses at primary (left) and secondary (right) minimums, respectively; the black points are the folded points, and the blue lines are the folded and binned points.\label{KIC011495766.eps}}
\end{figure*}

The object KIC 11495766 was classified to be a planet candidate with one planet of 4.175 d period by Burke et al. (2014) and Rowe et al. (2015), so it was also named KOI-2630. However, it is suggested that the KIC 11495766 is a triple system containing a stellar binary of $\sim120.73$ d period, and very close to it in celestial sphere there is a blended background stellar binary system with 8.3404432 d period. The background object is close enough to contaminate the light curve of KIC 11495766 in some observation quarters (one quarter is about three months in \textit{Kepler} mission), because the \textit{Kepler} spacecraft re-orientates itself each quarter, changing the contamination of the target, so the background object will sometimes be measured together with KIC 11495766 and sometimes not. As a result of this problem, the light curve of the background binary will show up in some quarters and disappear in others. Specifically, the 8.3404432 d period eclipses only appear in one quarter among the 15 quarters (see the blue bar in the upper panel of Figure \ref{KIC011495766.eps}). Therefore, the 8.3404432 d period eclipses should come from a blended background source, but not from KIC 11495766 itself.

The phase curves of the 8.3404432 d period around eclipses are shown in the lower panels of Figure \ref{KIC011495766.eps}. It can be seen that the difference between the primary and secondary eclipse is visible but not obvious, and the position of the secondary minimum (see the lower right panel) is clearly deviated from 0.5 phase. These indicate that the 8.3404432 d period eclipses are caused by a stellar binary system in a slightly eccentric orbit.

\subsubsection{The extraneous eclipses of $\sim120.73$ d period}

\begin{figure*}
\centering
\includegraphics[scale=.30]{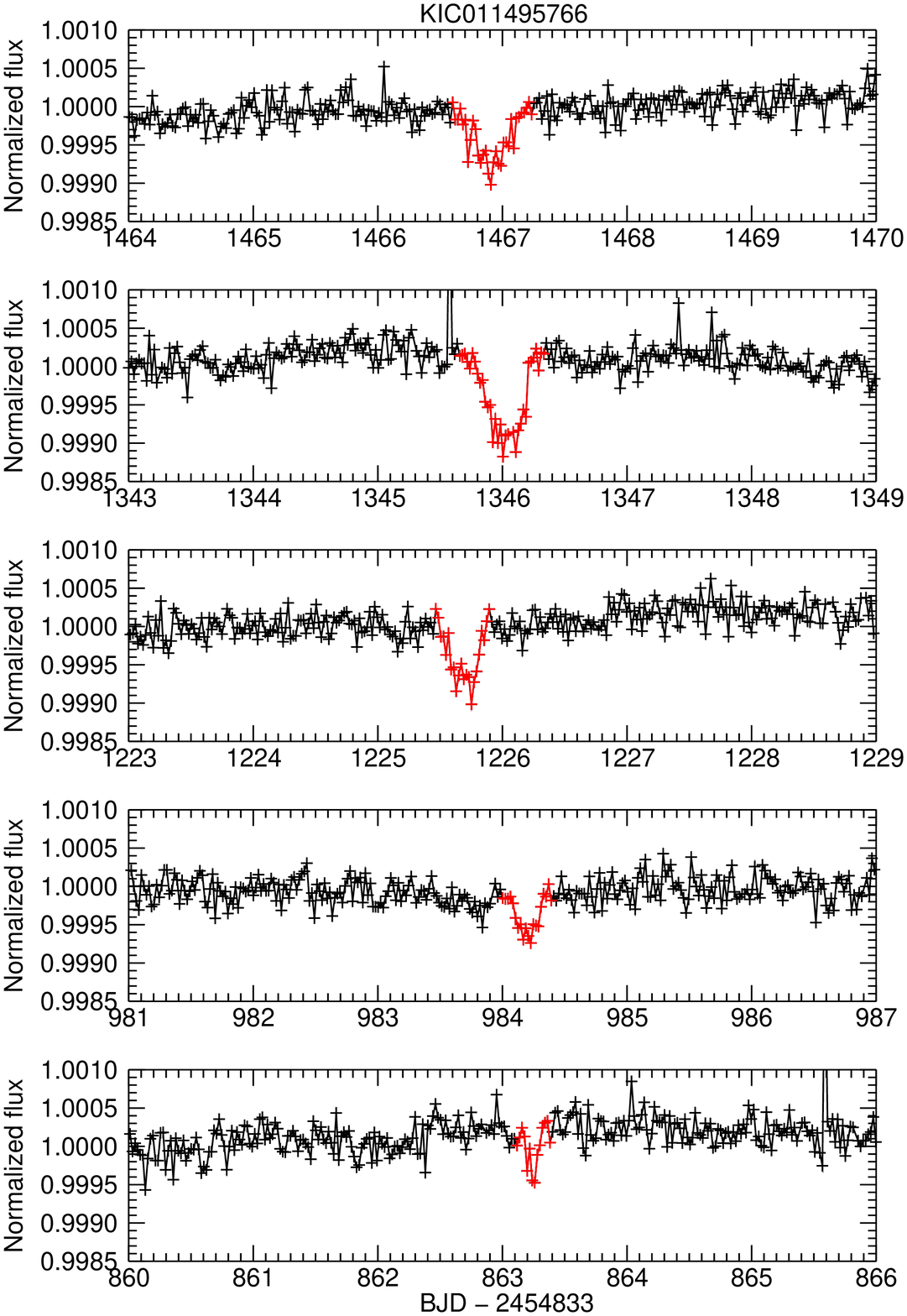}
\includegraphics[scale=.30]{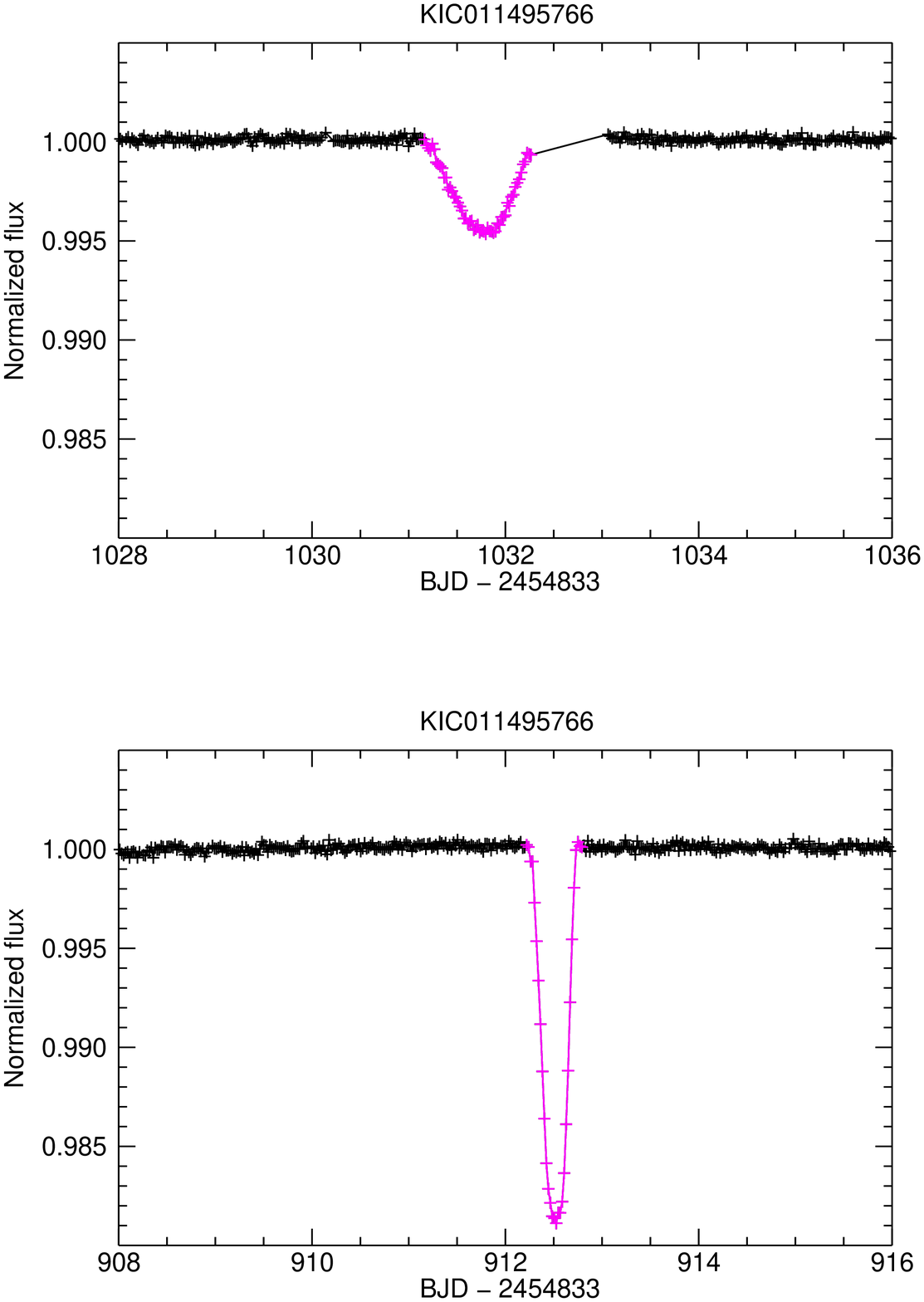}
\caption{The extraneous eclipses of KIC 11495766. Left panels: the red points are the extraneous secondary eclipses. The time intervals between the red eclipses are, or times of, $\sim120.73$ d. Right panels: the magenta points are the extraneous primary eclipses with a time interval of 119.3 d. \label{KIC011495766_extra_eclipses_1_and_2.eps}}
\end{figure*}

Besides the 8.3404432 d period eclipses, seven extraneous eclipses were found and shown in Figures \ref{KIC011495766.eps} and \ref{KIC011495766_extra_eclipses_1_and_2.eps}. The red and magenta points in the upper panel of Figure \ref{KIC011495766.eps} represent the extraneous eclipses, which can be seen in detail in Figure \ref{KIC011495766_extra_eclipses_1_and_2.eps}. The time intervals between the red eclipses (and equivalently the magenta eclipses) are approximately the same, that is, $\sim120.73$ d. The folded phase curves by $120.73$ d period\footnote{The phase curves were calculated only from the data larger than 800 d where the red eclipses exists.} were shown in the middle panel of Figure \ref{KIC011495766.eps}, where all the red eclipses located at phase 1 and the magenta eclipses located around phase 0.4. The existence of both primary and secondary eclipses with sharp shape suggests a stellar binary system rather than a planet system, because in a planet system the secondary eclipse is often flat and almost invisible.

For both the left panels (red eclipses) and the right panels (magenta eclipses) in Figure \ref{KIC011495766_extra_eclipses_1_and_2.eps}, the lengths of the abscissa and ordinate are the same. Therefore, we can easily estimate the changes of the width and depth of the eclipses by eye. For the red eclipses, they become wider and deeper with time (from bottom to top), and become shallower and narrower at the last eclipse. For the magenta eclipses, they change to wider and shallower by several times. In addition to the changes in the eclipse shape, the changes in the periods are also dramatic. The periods, that is, the time intervals between two successive eclipses, change from 120.95 d down to 120.33 d and then up to 120.93 d for the red eclipses. Meanwhile, the time interval of the two magenta eclipses is 119.3 d, that is, 1.65 d smaller than the period 120.95 d of the nearby red eclipses. Furthermore, the two striking magenta eclipses appear and disappear suddenly after and before (at least) several cycles.

The reason for such dramatic and rapid changes could not have come from the binary itself because the changes driven by the internal factors within a binary system, such as apsidal motion, stellar evolution,  and so on, are very small and take part very slowly. The eclipses and periods could never change visibly within a time span of one to two orbital periods. Therefore, the dramatic changes in the eclipses hint at strong disturbance force from (at least) an extraneous body. That is to say, the $\sim120.73$ d period binary must belong to a triple system; a simple binary system could not change in the manner shown in Figure \ref{KIC011495766_extra_eclipses_1_and_2.eps}.

Further information can be gained about the $\sim120.73$ d period binary through the comparison between the red and magenta eclipses. Because the red eclipses are narrower and shallower than the magenta eclipses, the red (magenta) eclipses should be caused by the cooler (hotter) component blocked by the hotter (cooler) one near the periastron (apastron). The changes of the red eclipses may be the result of the movement of the eclipse point toward the periastron, and meanwhile the magenta eclipse point moves toward the apastron causing it to become wider and shallower. Since the red and magenta eclipses have large differences in depth (10 to 40 times difference), the two components should have greatly different temperatures (perhaps an approx. two-fold difference, because the fourth power of temperature ratio is proportional to the luminosity ratio). Furthermore, the difference in mass is probably large (likely more than two-fold difference, considering the stellar mass luminosity relation).

It is suggested that the $\sim120.73$ d period binary consists of a star and a much less massive component, and around them there is another body with strong gravitational perturbation on the binary. Therefore, the object KIC 11495766 is suggested to be a triple system containing an eclipsing stellar binary with rapid changes. It should be noted that very close to KIC 11495766 in celestial sphere, there is a background object of another stellar binary system with an 8.3404432 d period.

\section{A first list of eclipsing multiple candidates}

A first list of eclipsing multiple candidates is presented in Table \ref{tab:objects_list}, and all the objects were found by the \textit{Kepler} data. For KIC 7941635, a dramatic huge extraneous fading event was found at 420 d from a version of data from August 1, 2014, which is not present in a later version of the data. The older version of the data has been shown, by the case of KIC 7668648 and KIC 8938628 (Zhang et al. 2017), to be better at reserving authentic extraneous eclipses than the newer version data. Therefore, KIC 7941635 is also listed in Table \ref{tab:objects_list}. The disappeared fading is probably caused by the improved PDC which works well for the majority of Kepler targets, but inevitably degrades performance for some individual targets by removing legitimate stellar signals on them.

All the periods of the inner binaries are taken from the catalog of \textit{Kepler} Eclipsing Binary Stars (Pr{\v s}a et al. 2011; Slawson et al. 2011; Matijevi{\v c} et al. 2012; Conroy et al. 2014; Kirk et al. 2016). The numbers of extraneous eclipses were counted from our processed data (except for KIC 9632895), and most of the period change amplitudes were estimated from the O-C curves calculated by us. Some O-C curves have obvious variation, and then a general change amplitude is given; some others are very diffuse and chaotic, so only a rough upper value headed with a $<$ symbol is given. For these latter curves, minimum time cannot be measured, and so neither for O-C values, change ranges are null in the table. The object KIC 11495766 is not an authentic eclipsing multiple system, so it is not listed here.

\begin{table*}[h!]
\small
\begin{center}
\begin{threeparttable}
\begin{minipage}[]{100mm}
\caption{A list of 25 eclipsing multiple candidates.\label{tab:objects_list}}\end{minipage}
\setlength{\tabcolsep}{1pt}
\begin{tabular}{llllll}
\hline
\hline
Name &  Inner binary  &  Number of     & Inner period      &  Structure & Reference \\
     &  period (d)    & extraneous eclipses & change range (d)\tnote{a}  &            &           \\
\hline
KIC  2835289  &   0.8577619 &    4          &    0.025  &   hierarchical       &   Conroy et al. (2015)     \\
KIC  2856960  &   0.2585073 &    $\sim$48   &    0.004 &    hierarchical       &  Marsh et al. (2014)    \\
KIC  4150611  &   1.5222786 &    23+160     &     --    &   hierarchical+binary     &   Shibahashi \& Kurtz (2012)   \\
KIC  4247791  &   4.0497388 &    649        &  $<$0.001   &   double binary   &    Lehmann et al. (2012)    \\  
KIC  4862625  &   20.000249 &    $\sim10$\tnote{c}  &   $<$0.001  &   circumbinary planet      &    Kostov et al. (2013)    \\
KIC  5255552  &   32.448635 &    10        &    0.3    &   hierarchical         &    Borkovits et al. (2015) \\
KIC  5897826  &   33.794718 &    68        &     --    &   hierarchical         &    Carter et al. (2011)  \\
KIC  5952403  &   0.9056778 &    $\sim$26   &    0.006  &   hierarchical        &   Derekas et al. (2011)    \\
KIC  6543674  &   2.3910305 &    2          &    0.0065  &   hierarchical       &    Masuda et al. (2015)   \\
KIC  6762829  &   18.795266 &    $\sim13$\tnote{c}  &   $<$0.005  &   circumbinary planet      &   Orosz et al. (2012)     \\
KIC  6964043  &   5.3626595 &    17         &   0.2     &   hierarchical          &   Borkovits et al. (2015)  \\
KIC  7289157  &   5.2664250 &   $\sim$12    &   0.02    &   hierarchical         &    Borkovits et al. (2015)  \\
KIC  7622486  &   2.2799960 &    31         &   $<$0.005  &   double binary   &       Zhang et al. (2017) \\
KIC  7668648  &   27.818590 &    10+5      &    0.13   &   hierarchical         &    Borkovits et al. (2015)  \\
KIC  7670485  &   8.4677064 &    1          &   $<$0.001  &   hierarchical        &   Zhang et al. (2017)     \\
KIC  7941635  &   0.7627242 &    1?         &    $<$0.002 & hierarchical\tnote{?}   &  This paper         \\
KIC  8572936  &   27.795808 &    $\sim5$\tnote{c}    &   $<$0.002  &   circumbinary planet      &    Welsh et al. (2012)   \\
KIC  8938628  &   6.8622157 &    6+4        &   0.01    &   hierarchical         &    Borkovits et al. (2015)  \\
KIC  9007918  &   1.3872069 &    1          &   0.0016  &   hierarchical         &    Borkovits et al. (2016)  \\
KIC  9632895  &   27.322046 &    $3$\tnote{b}    &    --     &   circumbinary planet       &    Welsh et al. (2015)  \\
KIC  9837578  &   20.733749 &    $\sim10$\tnote{c}  &   $<$0.001  &   circumbinary planet      &    Welsh et al. (2012)    \\
KIC 10020423  &   7.4483776 &    $\sim28+4$\tnote{c} &  $<$0.002  &   circumbinary planet      &   Orosz et al. (2012b)   \\
KIC 10091110  &   4.2185174 &    $\sim$89   &   $<$0.002   &   double binary     &    This paper    \\
KIC 12351927  &   10.116148 &    3+1        &   $<$0.003  &   circumbinary planet      &   Kostov et al. (2014)     \\
KIC 12644769  &   41.077587 &    7         &   $<$0.003  &   circumbinary planet      &    Marsh et al. (2014)   \\
\hline
\hline
\end{tabular}
\begin{tablenotes}
  \item[a] All the numbers with $<$ are the estimated upper limit. The value is the whole variation range of the $O-C$ curve. If the variation is sinusoidal, this value should be twice the amplitude of the sine curve.
  \item[b] Welsh et al. (2015).
  \item[c] Estimated from the outer planet period.
\end{tablenotes}
\end{threeparttable}
\end{center}
\end{table*}

\section{Summary and discussion}

In this paper, three \textit{Kepler} binaries with extraneous eclipses on the binary light curves were studied. The object KIC 5255552 is an eclipsing triple system with a  rapidly changing inner binary and a hierarchical tertiary body in $862.1(\pm0.1)$ d period. The tertiary body is confirmed by both the extraneous eclipses and the $O-C$ analysis by Borkovits et al. (2016). Four Segmented inner binary light curves were analyzed independently, so the parameter changes with time can be seen quantitatively. A small suspicious eclipse at 1279 d was found, suggesting an additional possible body, making KIC 5255552 a potential quadruple eclipsing system.

The object KIC 10091110 is suggested to be a quadruple stellar system consisting of two eclipsing binaries with coincidental and interesting features. Based on the light curve analysis using a binary model,  both binaries have similar extremely small mass ratios ($0.060(13)$ and $0.0564(18)$) and similar mean primary densities ($0.3264(42)\;\rho_\odot$ and $0.3019(28)\;\rho_\odot$). Not only that, the orbital period ratio of the two binaries is very close to integer 2 (8.5303353/4.2185174 = 2.022). This could easily cause one to consider orbital resonance, but this is in conflict with the lack of ETV. We wondered if these coincidences suggest some physical relationship between the two pairs of binaries. Follow-up observations, especially of the high-resolution spectra, are suggested to reveal the mystery of KIC 10091110.

Excluding the already known 8.3404432 d period eclipses on KIC 11495766, seven extraneous eclipses with a period of $\sim120.73$ d were discovered. In this paper, the $\sim120.73$ d period eclipses are interpreted as being generated by KIC 11495766, and the 8.3404432 d period eclipses were deemed to be caused by a blended background source near KIC 11495766. It cannot be overlooked that the $\sim120.73$ d eclipses exhibit fast and dramatic changes which cannot be explained by the physics of binary itself. Therefore, there must be additional bodies responsible for the eclipse changes, and the object KIC 11495766 must be (at least) a triple system, despite lacking outer eclipses.


A list of 25 eclipsing multiple candidates is presented for the first time. All the 25 eclipsing multiple candidates come from the \textit{Kepler} observations, and they are perhaps the most reliable candidates so far. The eclipsing multiple systems are not only the multiple systems, their extraneous eclipses provide a new important resource for research on multiple systems. This list could be greatly expanded by future projects similar to the \textit{\textcolor[rgb]{0,0,0}{K}epler} survey, and we hope this first small list could be beneficial for studying eclipsing multiples.

\begin{acknowledgements}
We thank the anonymous referee for constructive criticisms and recommendations which have significantly improved this manuscript. This paper includes data collected by the \textit{Kepler} mission. Funding for the \textit{Kepler} mission is provided by the NASA Science Mission directorate.
We are indebted to the \textit{Kepler} Science Working Group on Eclipsing Binaries for all their great pilot works which made these discoveries
possible. This work is partly supported by Chinese Natural Science Foundation (No. 11703082, 11133007, 11325315, 11403056), the West Light Foundation of The Chinese Academy of Sciences, Yunnan Natural Science Foundation (Y5XB071001), Guangdong Provincial Engineering Technology Research Center for Data Science, the Joint Research Fund in Astronomy (grant number U1631108) under cooperative agreement between the National Natural Science Foundation of China (NSFC) and Chinese Academy of Sciences (CAS), and the research fund of Sichuan University of Science and Engineering (grant number 2015RC42), the Key Research Program of the Chinese Academy of Sciences (grant No. KGZD-EW-603), the Science Foundation of Yunnan Province (No. 2012HC011), and the Strategic Priority Research Program ``The Emergence of Cosmological Structures'' of the Chinese Academy of Sciences (No. XDB09010202).
\end{acknowledgements}

%
%

\end{CJK*}
\end{document}